\documentclass[aps,amsmath,amssymb,preprintnumbers,nofootinbib,a4paper,prl,twocolumn]{revtex4-1}
\pdfoutput=1
\usepackage{amsthm}
\usepackage{graphicx}
\usepackage{color}
\usepackage{bbm}
\usepackage{pxfonts}
\def\beq{\begin{equation}}
\def\eeq{\end{equation}}
\def\be{\begin{equation}}
\def\ee{\end{equation}}
\def\bea{\begin{eqnarray}}
\def\eea{\end{eqnarray}}
\def\ba{\begin{eqnarray}}
\def\ea{\end{eqnarray}}
\usepackage{geometry}

\newcommand{\ft}{{\cal F}_t}
\renewcommand{\(}{\left(}
\renewcommand{\)}{\right)}

\usepackage{epsfig}
\usepackage{multirow}

\usepackage[margin=5pt, font=normalsize,labelfont=bf,justification=raggedright]{caption} 

\definecolor{rossoCP3}{cmyk}{0,.88,.77,.40}
 
\begin{document}

\title{\Large  \color{rossoCP3}  Large tensor-to-scalar ratio from Composite Inflation} 
\author{Phongpichit Channuie}
\email{phongpichit.ch@wu.ac.th} 

\affiliation{
\vspace{5mm} 
{Physics Division, School of Science, Walailak University, \\Nakhon Si Thammarat 80160, Thailand}}

\author{Khamphee Karwan}
\email{khampheek@nu.ac.th} 

\affiliation{
\vspace{5mm} 
{The Institute for Fundamental Study, Naresuan University,
\\Phitsanulok 65000 and Thailand Center of Excellence in Physics, \\Ministry of Education,
Bangkok 10400, Thailand}}

 \begin{abstract}
 The claimed detection of the BICEP2 experiment on the primordial B-mode of cosmic microwave background polarization suggests that cosmic inflation possibly takes place at the energy around the grand unified theory scale given a constraint on the tensor-to-scalar ratio. i.e., $r\simeq 0.20$. In this report, we revisit single-field (slow-roll) composite inflation and show that, with the proper choice of parameters and sizeable number of e-foldings, a large tensor-to-scalar ratio consistent with the recent BICEP2 results can be significantly produced with regard to the composite paradigms. 
 \end{abstract}

\maketitle
 
Elucidating the underlying theory of the inflationary Universe is a central problem in cosmology. During the very early moments in the evolution of the Universe, scientists believe that the gravitational-wave background would have left an imprint on the polarization of the cosmic microwave background (CMB). Among interesting features, the primordial B- and E-mode polarizations have gained much attraction. The very recent announcement of the BICEP2 experiment \cite{Ade:2014xna} addresses the detection of the primordial B-mode polarization of the CMB originated from gravitational waves created by inflation, thereby giving us a strong support for the inflationary scenario \cite{Starobinsky:1979ty, Starobinsky:1980te, Mukhanov:1981xt, Guth:1980zm, Linde:1981mu,Albrecht:1982wi} taking place at the energy around $10^{16}$\,GeV, i.e. the grand unified theory (GUT) scale.

A series of papers on model updates has been resurrecting by the exciting new results of the BICEP2 experiment. Theses recent investigations include the Higgs-related inflationary scenarios \cite{Nakayama:2014koa,Cook:2014dga,Hamada:2014iga,Germani:2014hqa,Oda:2014rpa}, several paradigms of chaotic inflation \cite{Harigaya:2014sua,Lee:2014spa}, some interesting analyses related to supersymmetry \cite{Harigaya:2014qza,Czerny:2014qqa}, and other compelling scenarios \cite{Ellis:2014cma,Viaggiu:2014moa,Kehagias:2014wza,Kobayashi:2014jga,Hertzberg:2014aha,Ferrara:2014ima,Gong:2014cqa,Okada:2014lxa,Bamba:2014jia,DiBari:2014oja,Palti:2014kza,Kumar:2014oka,Fujita:2014iaa,Chung:2014woa,Antusch:2014cpa,Bastero-Gil:2014oga,Kawai:2014doa,Hossain:2014coa,Kannike:2014mia,Ho:2014xza}. BICEP2's detection of the B-mode power spectrum constrains the ratio of tensor-to-scalar perturbations to be $r = 0.20^{+0.07}_{-0.05}$ at $68\%$ C.L. with no foreground subtraction. This raises the possibility to look for the signal of gravitational waves possibly generated during inflation. The possibility that the ratio is zero is ruled out with a statistical certainty of $7\sigma$. 

However, it is considered to be significantly larger than that expected from previous results of Planck and WMAP \cite{Planck}. This apparent tension can be alleviated in accordance with recent various investigations, e.g. running of the spectral index \cite{Choudhury:2014kma,Czerny:2014wua,Garcia-Bellido:2014gna} and the presence of sizeable quantum departures from the $\phi^{4}$-inflationary model with the nonminimally coupled scenario \cite{Joergensen:2014rya}. De facto, however, all of the above investigations deal with inflation driven by an elementary inflaton field.

\begin{figure}
\begin{center}
\includegraphics[width=3.0in]{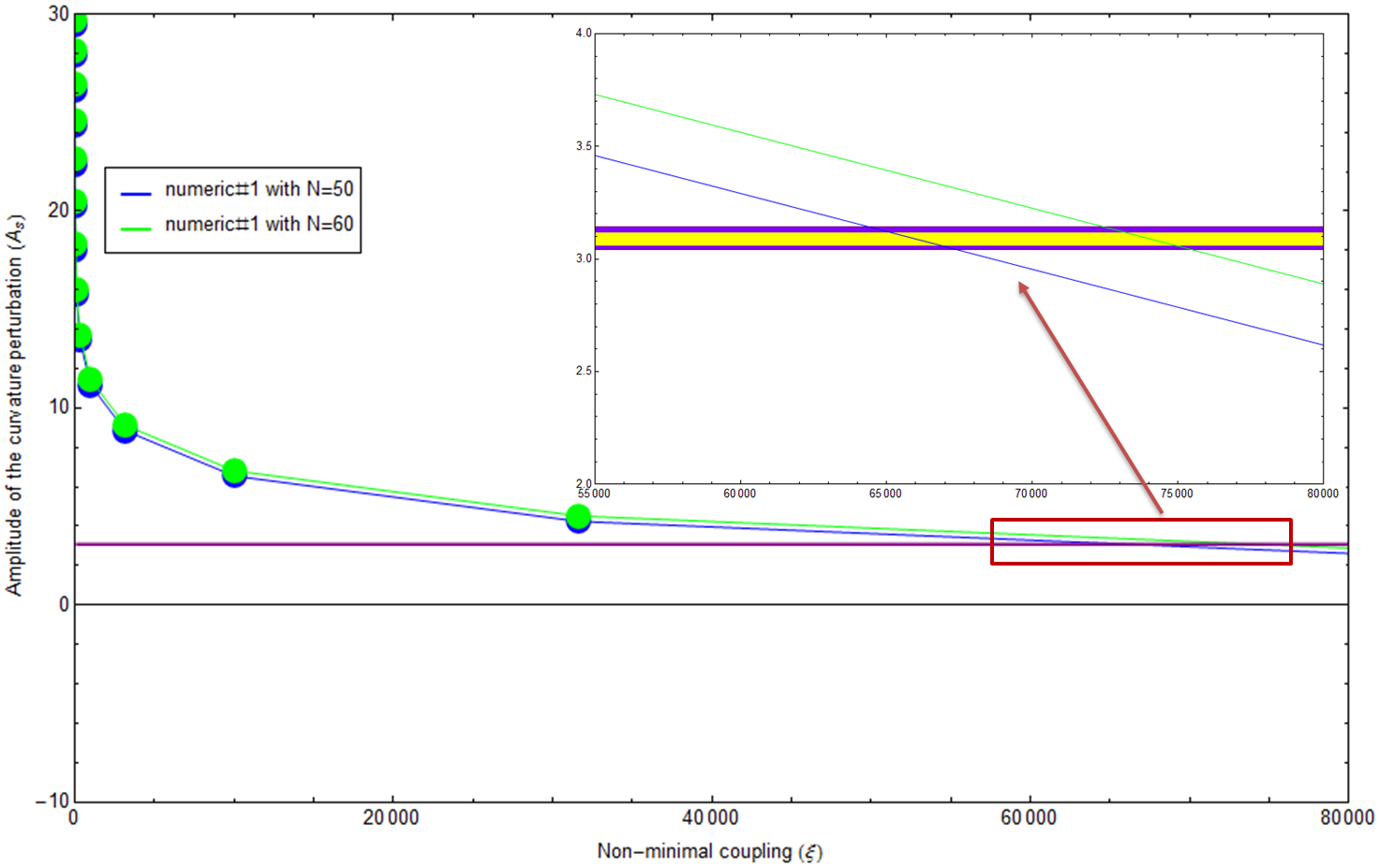} 
\end{center}
\caption{The plot shows the relation between the amplitude of the power spectrum ${\cal A}_{s}$ and the nonminimal coupling $\xi$ with $10^{-3}\lesssim \xi\lesssim 10^{6}$ for ${\cal N}=50,60$ predicted by model$1$. The horizontal bands represent the $1\sigma$ (yellow) and $2\sigma$ (purple) C.L. for ${\cal A}_{s}$ obtained from Planck.} 
\label{f1gb}
\end{figure}

\begin{figure}
\begin{center}
\includegraphics[width=3.0in]{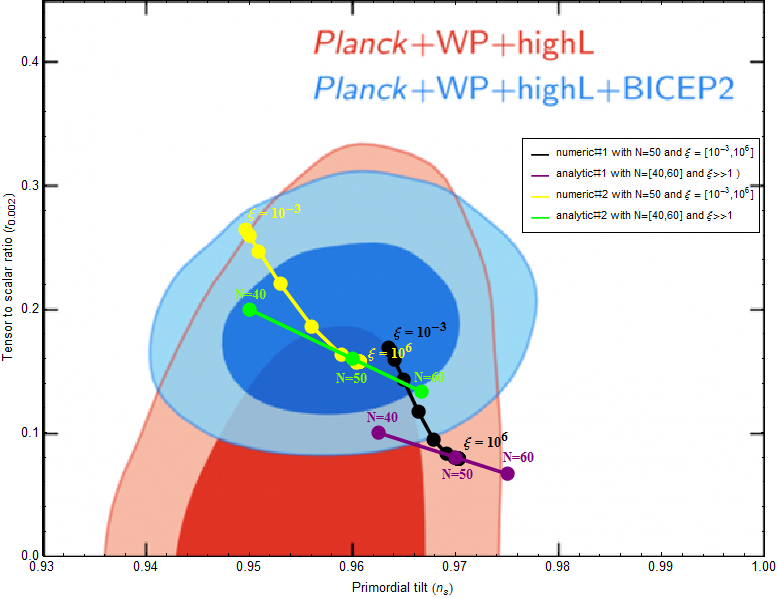} 
\end{center}
\caption{The contours show the resulting 68$\%$ and 95$\%$ confidence regions for the tensor-to-scalar ratio $r$ and the scalar spectral index $n_{s}$. The red contours are for the Planck+WP+highL data combination, while the blue ones display the BICEP2 constraints on $r$. The plots show the analytical and numerical data predicted by the models examined in this work ($\#1$ and $\#2$).} 
\label{f1to}
\end{figure}

The authors of Refs. \cite{Evans:2010tf,Channuie:2011rq,Bezrukov:2011mv,Channuie:2012bv,Channuie:2013lla,Channuie:2013xoa} have posted the compelling assumption that the inflaton needs not be an elementary degree of freedom called the \lq\lq composite inflaton\rq\rq and remarkably showed that the energy scale of inflation driven by composite inflaton is around the GUT energy scale \cite{Channuie:2011rq,Bezrukov:2011mv,Channuie:2012bv}. We will show in this report that composite paradigms potentially provide the tensor-to-scalar ratio $r$, consistent with BICEP2 observations, and the scalar spectral index $n_{s}$ and the amplitude of scalar perturbation ${\cal A}_{s}$, consistent with Planck results by using the proper choice of parameters and sizeable number of e-foldings. The general action for composite inflation in the Jordan frame takes the form for scalar-tensor theory of gravity as
\begin{eqnarray}
\mathcal{S}_{\rm CI,J}=\int d^{4}x \sqrt{-g}&~&\Big[\frac{M^{2}_{\rm P}}{2} F(\Phi) R - \frac{1}{2}G(\Phi)g^{\mu\nu} \partial_{\mu}\Phi\partial_{\nu}\Phi \nonumber\\&&- V(\Phi)  \Big]. \label{action}
\end{eqnarray}
Here $F(\Phi)$ and $G(\Phi)$ in this action are functions of the field $\Phi$ and can be written as
\begin{equation}
F(\Phi) = 1 + \frac\xi{M^{2}_{\rm P}}\,\Phi^{2/D} \,\, {\rm and} \,\, G(\Phi) = \frac{1}{D^2}G_{0}\Phi^{(2 - 2D)/D}\,,
\label{fg1}
\end{equation}
where $D$ is the mass dimension of the composite field $\Phi$, $G_{0}$ is a constant, 
and $1/D^2$ is introduced for later simplification.
In this investigation, we write the potential in the form
\begin{equation}
V(\Phi) = \Phi^{4/D} f(\Phi)\quad\quad{\rm with}\quad\quad\Phi \equiv \varphi^{D} \,,
\label{fg}
\end{equation}
where the field $\varphi$ possesses a unity canonical dimension and $f(\Phi)$ is a general function of the field $\Phi$ concretely implemented below. The nonminimal coupling to gravity is controlled by the dimensionless coupling $\xi$.
Here, we write the general action for the composite inflation in the form of scalar-tensor theory of gravity in which the inflaton non-minimally couples to gravity.
At first glance, the nonminimal term $\xi\Phi^{2/D}R/M^{2}_{\rm P}$ has purely phenomenological origin. The reason resides from the fact that one wants to relax the unacceptable large amplitude of the primordial power spectrum if one takes $\xi=0$ or smaller than ${\cal O}(10^{4})$. With the nonminimal coupling term phenomenologically introduced, it is more convenient to diagonalize into another form by applying a conformal transformation. To this end, we take the following replacement:
\be
g_{\mu\nu} \longrightarrow \tilde{g}_{\mu\nu} = F\(\Phi\)g_{\mu\nu}\,.
\ee
With the above replacement, the action in Eq.~(\ref{action}) can be transformed into the new frame -- the Einstein frame -- as
\be
\mathcal{S}_{\rm CI,E}=\int d^{4}x \sqrt{-\tilde{g}}\Big[\frac{M^{2}_{\rm P}}{2} \tilde{R} - \frac{1}{2}\partial_{\mu}\chi\partial^{\mu}\chi - U(\chi)  \Big], 
\label{action-e}
\ee
where $\tilde g$ and $\tilde R$ are basically computed from $\tilde{g}_{\mu\nu}$; \lq\lq tildes\rq\rq\,represent the quantities in the Einstein frame, and
\be
\frac{\partial \Phi}{\partial \chi} = \frac F{\sqrt{G F + 3 F_{\Phi}^2 / 2}}\,
\,\,{\rm and}\,\,
U(\chi) = \left. \frac{V(\Phi)}{F^2(\Phi)} \right|_{\Phi = \Phi(\chi)}\,,
\ee
where the subscript denotes a derivative with respect to $\Phi$. We can reexpress inflationary parameters and all relevant quantities in terms of the field $\chi$ if we solve
\be
\chi \equiv \int \frac{\sqrt{G F + 3 F_{\Phi}^2 / 2}}{F}d\Phi\,.
\ee
Using the expression for the slow-roll parameter, $\tilde\epsilon$, in the Einstein frame such that
\be
\tilde\epsilon = \frac 12 \(\frac 1{U}\frac{\partial U}{\partial \chi}\)^2\,,
\label{epsilon-e}
\ee
one can simply show that
\be
\tilde\epsilon = \frac 12 \(\frac{F^2}{V}\frac{\partial \Phi}{\partial\chi}\frac{\partial }{\partial \Phi}\(\frac{V}{F^2}\)\)^2
= \epsilon + \ft\,,
\label{eps-e-eps-j}
\ee
where $\ft \equiv {\dot F}/2HF$; $\epsilon$ is the slow-roll parameter in the Jordan frame given by $\epsilon \equiv \ft - (V_{\Phi}/V)(F/F_{\Phi})\ft$; and the dot denotes a derivative with respect to time, $t$. It is well known that the power spectrum for the scalar perturbation generated from inflaton field $\chi$ in the Einstein frame is given by
\be
{\cal P}_{\zeta} \simeq  \left.\frac{U}{24\pi^2 \tilde\epsilon}\right |_{k |\tau| = 1}\,,
\label{pr-e}
\ee
where the above expression is evaluated at the conformal time $\tau$ when the perturbation with wave number $k$ exits the horizon and the tensor-to-scalar ratio is
\begin{eqnarray}
r  \simeq16\tilde\epsilon\,.
\label{t2s-e}
\end{eqnarray}
Since the power spectra are frame independent, we can use Eq.(\ref{eps-e-eps-j}) to write the power spectrum in Eq.(\ref{pr-e})
and the tensor-to-scalar ratio in Eq.(\ref{t2s-e})
in terms of the Jordan frame parameters as
\ba
{\cal P}_{\zeta} &\simeq& \left.\frac{V}{24\pi^2 F\Big(\epsilon + {\cal F}_{t}\Big)}\right |_{k |\tau| = 1}\,,
\label{pr-cal}\\
r &\simeq&16\left(\epsilon+{\cal F}_{t}\right)\,.
\label{t2s}
\ea
Here, it is convenient (although tricky) to use the results in the Einstein frame, and then we transform the quantities in the Einstein frame into the Jordan one. It is noticed that one obtains the relation between two frames: ${\tilde \epsilon} \Leftrightarrow \epsilon+{\cal F}_{t}$. Having computed the field $\Phi$ at the end of inflation $\Phi_e$  by using the condition $\epsilon(\Phi_{e})=1$, one can determine the number of e-foldings via
\begin{eqnarray}
{\cal N}(\Phi) = \int_{\Phi}^{\Phi_e}\frac{H}{\dot{{\tilde \Phi}}}d{\tilde\Phi}=\int_{\Phi}^{\Phi_e}\frac 1{{\tilde \Phi}'}d{\tilde\Phi}\,,
\label{efold}
\end{eqnarray}
where the subscript \lq\lq $e$\rq\rq\,denotes the evaluation at the end of inflation
and $\Phi'$ is given by 
\begin{eqnarray}
\Phi'= \frac{1}{\left(1+ \frac{3F_{\Phi}^2}{2 F G}\right)}\Big(2 \frac{F_{\Phi}}{G} - \frac{V_{\Phi}}{V} \frac{F}{G}\Big)\,.
\label{epsi-ss}
\end{eqnarray}
Here, we have used the Friedmann equation and the evolution equations for the background field and apply the standard slow-roll approximations. 
Determining the value of $\Phi$ and $\Phi'$ when the perturbations exit the horizon allows us to compute the spectral index and the amplitude of the power spectrum in terms of the number of e-foldings. The spectrum index for this power spectrum can be computed via
\be
n_s = \frac{d\ln {\cal P}_{\zeta}}{d \ln k} + 1 \simeq 1 - 2\epsilon - 2 {\cal F}_{t}
- \Phi'\frac{d\ln(\epsilon+{\cal F}_{t})}{d\Phi}\,.
\label{ns}
\ee
The amplitude of the curvature perturbation can be directly read from the power spectrum and we find
\begin{eqnarray}
{\cal A}_{s}\equiv\log \left[|\zeta|^2\times 10^{10}\right] \simeq \log \left[\left.\frac{V\times10^{10}}{24\pi^{2} F^2\Big(\epsilon + {\cal F}_{t}\Big)}\right.\right]_{c_{s} k|\tau| = 1}\,.
\label{zeta2}
\end{eqnarray}
We consider the first viable model of composite inflation (model$1$) in which inflation is driven by gluonic-type fields. In this case, the inflaton emerges as the interpolating field describing the lightest glueball associated to a pure Yang--Mills theory. It is worthy to note here that the theory we are using describes the ground state of pure Yang--Mills theory, and of course is not the simple $\phi^4$ theory. For this model, we have
\begin{eqnarray}
 f(\varphi)=2\ln(\varphi/\Lambda)\,,
 \end{eqnarray}
so that the effective Lagrangian for the lightest glueball state, constrained by the Yang--Mills trace anomaly, nonminimally coupled to gravity in the Jordan frame reads
\begin{eqnarray}
{\cal S}_{\#1} = \int d^{4}x\sqrt{-g}&~&\Big[\frac{F(\varphi)}{2} R - 16g^{\mu\nu} \partial_{\mu}\varphi\partial_{\nu}\varphi \nonumber\\&&- V_{\#1}(\varphi)\Big]\,,
\end{eqnarray}
where
\begin{eqnarray}
V_{\#1}(\varphi) = 2\varphi^{4}\ln\left(\varphi/\Lambda\right)\,.
\label{v-gb}
\end{eqnarray}
In this work, we consider only the large $\xi$ limit and find for this case
\ba
{\cal N} \simeq 3\Big(\ln^2\(\varphi/\Lambda\) - \ln^2\(\varphi_{e}/\Lambda\)\Big) + {\cal O}(1/\xi)\,.
\ea
Here, we can write $\varphi$ in terms of ${\cal N}$ and use Eqs.(\ref{ns}), (\ref{t2s}), and (\ref{zeta2}) to write $n_s$, $r$, and $|\zeta|^2$ in terms of ${\cal N}$. Finally, we obtain for a large $\xi$ limit
\begin{eqnarray}
n_s &\simeq&  1-\frac{3}{2{\cal N}} + {\cal O}(\xi)\,,
\label{ns-gb-l}\\
r &\simeq&  \frac{4}{{\cal N}} + {\cal O}(\xi)\,,
\label{t2s-gb-l}\\
|\zeta|^2 &\simeq&  \frac{{\cal N}^{3/2}}{3\sqrt{3}\pi^2\xi^2} + {\cal O}(1/\xi^3)\,.
\label{zeta2-gb-l}
\end{eqnarray}
Notice that the above relations lead to the consistency relation, allowing us to write
\begin{eqnarray}
r \simeq  \frac{8}{3}(1-n_{s})\,.
\end{eqnarray}
From the above analytical estimations, we see that when $\xi \gg 1$, $n_s$ , $r$, and $|\zeta|^2$ can satisfy the 95$\%$ C.L. observational bound from Planck data for $50 < {\cal N} < 60$ and $\xi \sim 10^{4}$; see Fig.\,\ref{f1gb} and \,\ref{f1to}. Nevertheless, for such range of ${\cal N}$, $r$ lies outside the $2\sigma$ C.L. with BICEP2 results shown in Fig.\,\ref{f1to}.
The value of $r$ will increase and then satisfy the bound from BICEP2 results when ${\cal N}\lesssim 45$. However, it is obvious that ${\cal N}$ is a model-dependent quantity. However, it is quite subtle if we have ${\cal N}\lesssim 45$ for models of inflation to be viable. This is so since, in order to solve the horizon problem, in the common formulation one frequently uses at least ${\cal N}\subset [50, 60]$. We anticipate this can be further verified by studying the reheating effect. The compatibility between our analytical and numerical results of this model is illustrated in Fig.\,\ref{f1to}. 

According to the second model of inflation (model$2$), the inflaton is designed to be the gluino-ball state in the super--Yang--Mills theory. For this model, we have
\begin{eqnarray}
 f(\varphi)=4\alpha N^{2}_{c}\ln^{2}(\varphi/\Lambda)\,.
 \end{eqnarray}
As it is always investigated in standard fashion, we take the scalar component part of the superglueball action and coupled it nonminimally to gravity. Focusing only on the modulus of the inflaton field and taking the next step in order to write the non-minimally coupled scalar component part of the superglueball action to gravity, the resulting action in the Jordan frame reads
\begin{eqnarray}
{\cal S}_{\#2} = \int d^{4}x\sqrt{-g}&~&\Big[\frac{F(\varphi)}{2} R - \frac{9N^2_{c}}{2\alpha}g^{\mu\nu} \partial_{\mu}\varphi\partial_{\nu}\varphi \nonumber\\&&- V_{\#2}(\varphi)  \Big]\,,
\end{eqnarray}
where
\begin{eqnarray}
V_{\#2}(\varphi) = 4\alpha N^{2}_{c}\varphi^{4}(\ln[\varphi/\Lambda])^{2}\,,
\end{eqnarray}
with $N_{c}$ a number of colors, and $\alpha$ a $N_{c}$-independent quantity. Using the similar approximations to those of the above consideration, the number of e-foldings for this inflation model in the large $\xi$ limit is approximately given by
\ba
{\cal N} \simeq \frac{3}{2}\Big(\ln^2\(\varphi/\Lambda\) - \ln^2\(\varphi_{e}/\Lambda\)\Big) + {\cal O}(1/\xi)\,.
\label{efold-sgb-l}
\ea
Regarding to the above relations between the number of e-foldings and $\varphi$,
we can write $n_s$, $r$ and $|\zeta|^2$ in terms of ${\cal N}$ for a large $\xi$ limit to yield
\begin{eqnarray}
n_{s} 
&\simeq&  1 - \frac 2{{\cal N}} + {\cal O}(1/\xi)\,,
\label{ns-sgb-l}\\
r 
&\simeq&  \frac 8{{\cal N}} + {\cal O}(1/\xi)\,,
\label{t2s-sgb-l}\\
|\zeta|^2 &\simeq&  \frac{2\alpha{\cal N}^{2}}{81N^{2}_{c}\pi^{2}\xi^{2}} + {\cal O}(1/\xi^3)\,.
\label{zeta2-sgb-l}
\end{eqnarray}
The consistency relation of the above relations reads
\begin{eqnarray}
r \simeq 4(1-n_{s})\,.
\end{eqnarray}
\begin{figure}
\begin{center}
\includegraphics[width=3.0in]{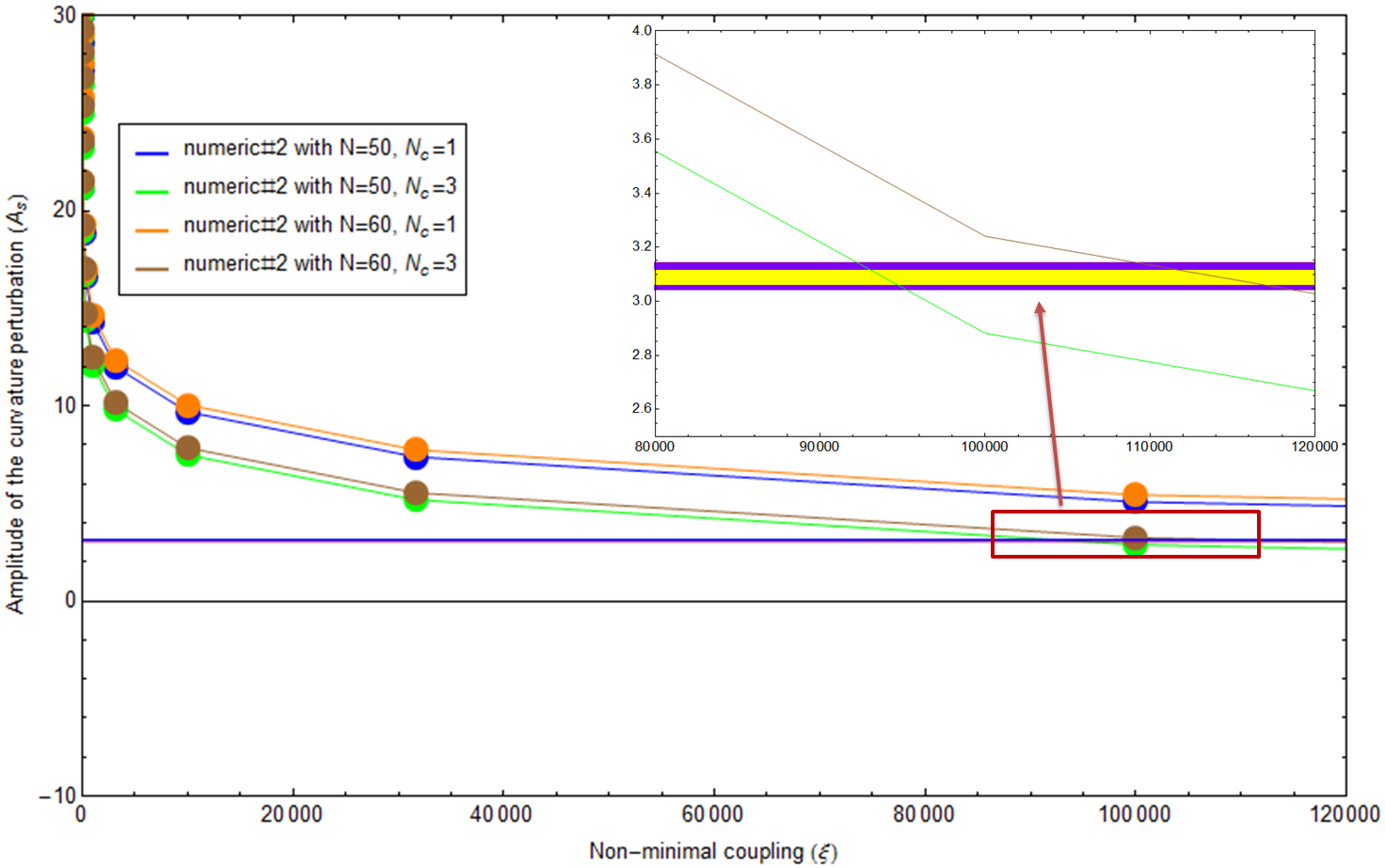} 
\end{center}
\caption{The plot shows the relation between the amplitude of the power spectrum ${\cal A}_{s}$ and the nonminimal coupling $\xi$ with $10^{-3}\lesssim \xi\lesssim 10^{6}$ for ${\cal N}=50,60$ predicted by model$2$. The horizontal bands represent the $1\sigma$ (yellow) and $2\sigma$ (purple) C.L. for ${\cal A}_{s}$ obtained from Planck.} 
\label{f2sgb}
\end{figure}
The main results from the above analytical estimations are similar to those of the preceeding ones. The interesting different result is that for this model of inflation $r$ can be large enough to satisfy the bound launched by BICEP2; see Fig.\,\ref{f1to}. Concretely, the predictions of this model are not only satisfactorily consistent with the BICEP2 data at $68\%$ C.L. if ${\cal N} \subset [40,\,60]$ and $\xi >> 1$ with an exceptional range of $\Lambda$ but also lie inside the $1\sigma$ C.L. of the Planck data for ${\cal N} \subset [45,\,60]$. The amplitude of the power spectrum ${\cal A}_{s}$ predicted by model$\#2$ is in good agreement with the Planck results, see Fig.\,\ref{f2sgb}, for the proper choice of parameters and sizeable number of e-foldings.

In the present work, we examine single-field (slow-roll) inflation in which the inflaton is a composite field stemming from two strongly interacting field theories. Here, one can effectively solve the cosmological \lq\lq hierarchy problem\rq\rq\, in the scalar sector of the inflation which is not solved by Higgs inflation. 
With the proper choice of parameters and sizeable number of e-foldings, we showed that the predictions are significantly consistent with the recent BICEP2 data given the tensor-to-scalar ratio $r$ of approximately $0.16$ as well as the spectral index of $0.96$ illustrated in Fig.\,\ref{f1to}. 
Particularly, the predictions from the model$2$ consistent with the PLANCK and BICEP2 observations can be inherently generated. The results shown in Fig.\,\ref{f1to} show that 
our models favour only the rather large value of the tensor-to-scalar ratio. This behavior is opposed to that present in the ordinary Higgs-like scenarios and provides the possibility to rule out the models if upcoming experiments detect a small value of $r$. However, our results help us make a strong support that the energy scale during inflation is at the GUT scale. According to the present investigations, the composite paradigms and their verifiable consequences, e.g. reheating mechanism \cite{Bezrukov:2008ut,GarciaBellido:2008ab,Watanabe:2006ku}, can possibly receive considerable attention for inflationary model buildings. However, comprehensive and thorough studies along the lines of compositeness are still required. We anticipate that the potential of upcoming experiments can shed light on (composite) inflation.

\vskip .4cm
\noindent
P.C. is financially supported by the Institute for the Promotion of Teaching Science and Technology (IPST) under the project of the \lq\lq Research Fund for DPST Graduate with First Placement\rq\rq\,with Grant No.\,033/2557. K.K. is supported by Thailand Research Fund (TRF) through Grant No. RSA5480009.

\end{document}